\documentclass[journal=jacsat,manuscript=article]{achemso}
\usepackage[version=3]{mhchem} 
\usepackage[T1]{fontenc}

\usepackage{graphicx}
\usepackage{natbib}
\usepackage{url}
\usepackage{textcase}
\usepackage{xcolor}
\usepackage{bm}
\usepackage[normalem]{ulem}

\def \bii{BiI$_{3}$} 
\def \Fref{Fig.~\ref}
\author{Selene Mor}
\affiliation{Department of Mathematics and Physics, Universit\`a Cattolica, I-25133 Brescia, Italy}
\altaffiliation{Interdisciplinary Laboratories for Advanced Materials Physics (I-LAMP), Universit\`a Cattolica, I-25133
Brescia, Italy}
\email{selene.mor@unicatt.it}

\author{Valentina Gosetti}
\affiliation{Department of Mathematics and Physics, Universit\`a Cattolica, I-25133 Brescia, Italy}
\altaffiliation{Interdisciplinary Laboratories for Advanced Materials Physics (I-LAMP), Universit\`a Cattolica, I-25133
Brescia, Italy}
\alsoaffiliation{Department of Materials Engineering, KU Leuven, Kasteelpark Arenberg 44, 3001 Leuven, Belgium}

\author{Vadim F. Agekyan}
\affiliation{St. Petersburg State University, St. Petersburg, 199034, Russia}

\author{Claudio Giannetti}
\affiliation{Department of Mathematics and Physics, Universit\`a Cattolica, I-25133 Brescia, Italy}
\altaffiliation{Interdisciplinary Laboratories for Advanced Materials Physics (I-LAMP), Universit\`a Cattolica, I-25133
Brescia, Italy}

\author{Luigi Sangaletti}
\affiliation{Department of Mathematics and Physics, Universit\`a Cattolica, I-25133 Brescia, Italy}
\altaffiliation{Interdisciplinary Laboratories for Advanced Materials Physics (I-LAMP), Universit\`a Cattolica, I-25133
Brescia, Italy}

\author{Stefania Pagliara}
\affiliation{Department of Mathematics and Physics, Universit\`a Cattolica, I-25133 Brescia, Italy}
\altaffiliation{Interdisciplinary Laboratories for Advanced Materials Physics (I-LAMP), Universit\`a Cattolica, I-25133
Brescia, Italy}

\title[]
  {Effect of photoinduced screening on the spectroscopic signature of exciton-phonon coupling}

\abbreviations{}

\begin{document}

\begin{tocentry}

\includegraphics[scale=0.86]{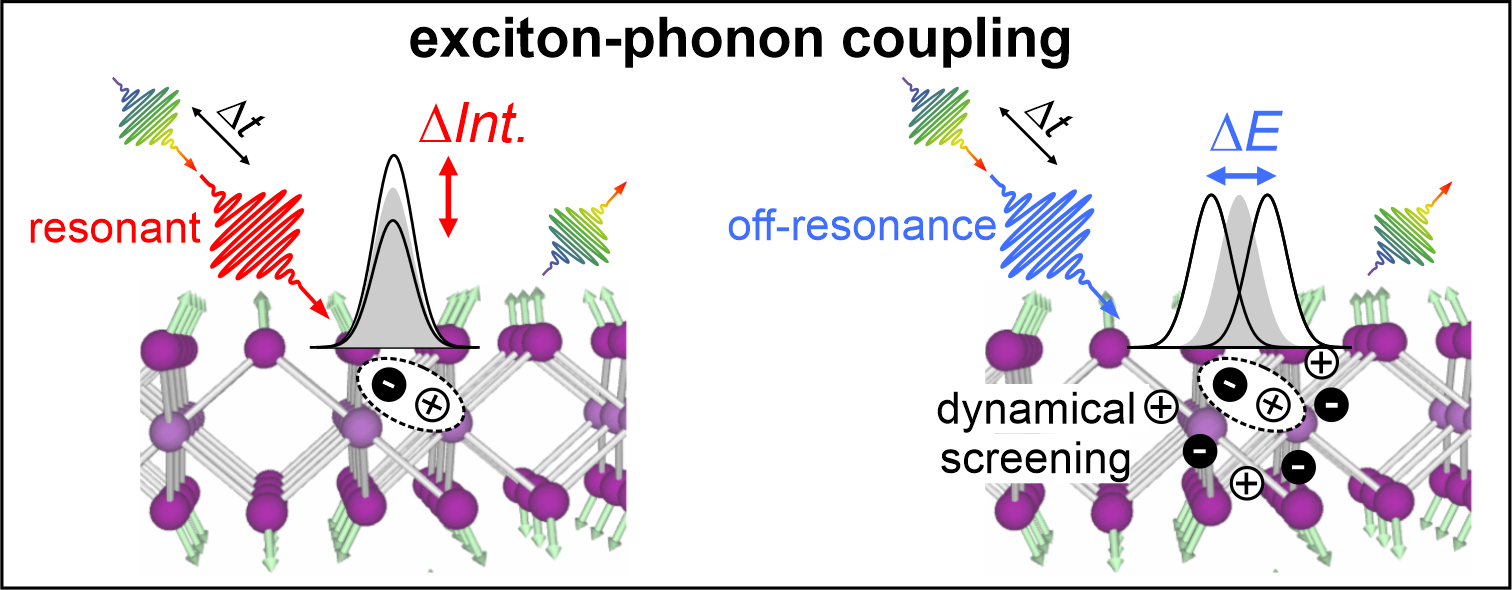}

\end{tocentry}

\begin{abstract}
 The light-mediated interaction of fermionic and bosonic excitations governs the optoelectronic properties of condensed matter systems. In photoexcited semiconductors, the coupling of electron-hole pairs (excitons) to coherent optical phonons enables a modulation of the excitonic resonance that is phase-locked to the frequency of the coupled vibrational mode. Moreover, due to the Coulombic nature of excitons, their dynamics are sensitive to transient changes in the screening by the photoexcited carriers. Interestingly, the effect of photoinduced screening on the transient optical signal originating from the exciton dynamics coupled to phonons is not yet established. By means of broadband transient reflectance spectroscopy, we disclose how exciton-phonon coupling manifests in either presence or absence of dynamical screening in a layered semiconductor. Further, we unveil the promoting role of photoinduced screening on these exciton-phonon coupled dynamics as opposed to the case in which the \textit{un}screened exciton-exciton repulsion likely dominates the nonequilibrium optical response. These findings set a protocol to look at an excitonic resonance and its fundamental many-body interactions on the ultrafast timescale, and provide new perspectives on the access to nonequilibrium coupled dynamics.

\end{abstract}

\section{Introduction}
Many-body interactions are at the heart of several paradigms of condensed matter physics~\cite{Bardeen1957, Mott1961, Jerome1967, Edwards1998, Imada1998}. Layered semiconductors formed by van-der-Waals stacked atomic planes offer an ideal platform for the study of these couplings: upon photoabsorption, spatially confined electron-hole pairs (excitons) are favored to couple with fermionic and bosonic excitations, such as quasi-free carriers and phonons, respectively~\cite{Shree2021, Batool2023}. In contrast to the bulk counterparts, where excitons are thermally unstable and rapidly recombine or remain trapped by defects~\cite{Teke2004, Foglia2019,Baldini2020}, the low density of mobile carriers of van-der-Waals semiconductors reduces the dielectric screening, thus enhancing the exciton binding energy up to several 10s meV at room temperature. As a result, a regime of excitons with well defined spectral lines~\cite{Chernikov2014} and dynamics up to the nanosecond timescale is reached upon photoexcitation even in the presence of quasi-free carriers. Remarkable examples are the nanoseconds (ns) lifetime of the valley polarization in transition metal dichalcogenide (TMDC) monolayers~\cite{DalConte2020}, and of the interlayer excitons in TMDC-based heterostructures~\cite{Rivera2015}, the  10s of picoseconds (ps) exciton relaxation dynamics in the van-der-Waals excitonic insulator candidate Ta$_2$NiSe$_5$ ~\cite{Mor2022, Katsumi2023}, up to the ns-long luminescence of layered perovskite-structured semiconductors, such as \bii~\cite{KAIFU1988} and MaPbBi$_3$~\cite{Fang2020}.

\begin{figure}
\includegraphics[width=0.7\columnwidth]{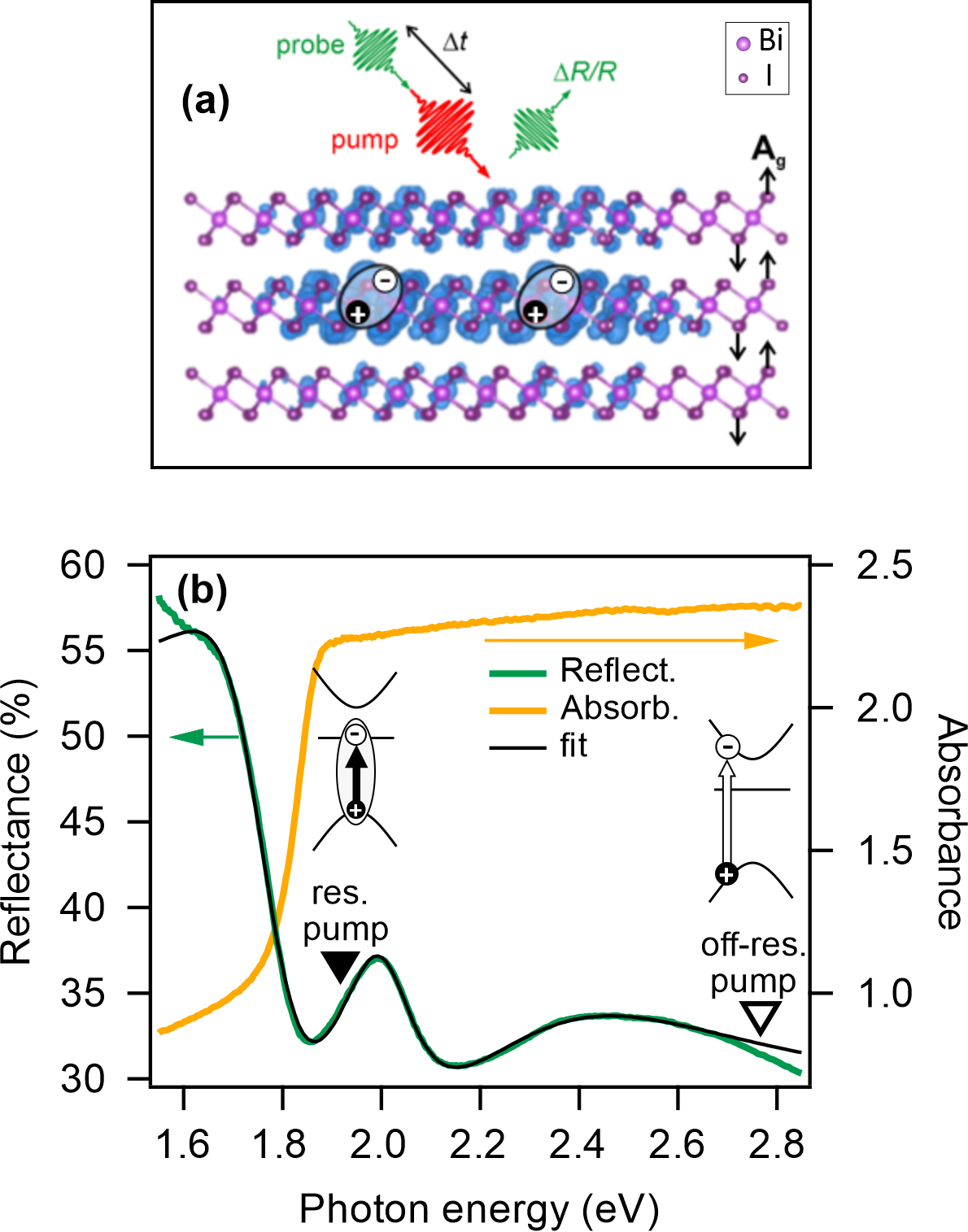}
\caption{(a) Schematic illustration of the TR experiment on a \bii\,single crystal. The real-space representation of two exciton wave functions (blue shades) are superimposed on the layered crystalline structure (modified from Mor \textit{et al.}~\cite{Mor2021}). The direction of the main $A_g$ Raman mode is indicated by the black arrows. (b) Steady-state reflectance (green) and absorbance (orange) spectra at room temperature. The black line is the Tauc-Lorenz fit to the reflectance spectrum. The black and white triangles indicate the resonant and off-resonance pump pulse energies, respectively, as used in the TR experiments.}
\label{Figure 1}
\end{figure}

Once excitons are formed, their dynamics evolve on the femtoseconds (fs) to several picoseconds (ps) timescale as governed by the exciton-exciton interaction and the exciton coupling to other quasiparticles like quasi-free carriers and phonons. Since these interactions and couplings are Coulombic in nature, we expect the screening to play a significant role. Particularly, \textit{transient} screening enhancement by the photoexcited charges is well known to renormalize (i.e. reduce) both the electronic band gap and the exciton binding energy~\cite{Reynolds2000, Pagliara2011, Nie2014, Pogna2016, Sangalli2015, Pogna2016, Mor2017, Cunningham2017, Baldini2018, Liu2019, Dendzik2020, Wang2020, Mor2022, Calati2023}, even more drastically in layered semiconductors than in bulk systems~\cite{Chernikov2014,Chernikov2015,Sie2017, Baldini2019}. In layered semiconductors, the in-plane localization of the exciton wave function (\Fref{Figure 1}(a)) favors strong dipole-dipole interaction among excitons. On the one hand, the attraction between two excitons often leads to the formation of biexciton bound states few 10s of meV  below the single exciton transition~\cite{Rodin2020}. On the other hand, when the exciton-exciton interaction becomes repulsive it manifests thorugh a transient blue-shift of the excitonic resonance~\cite{Shahnazaryan17, Erkensten21}. Additionally, the coupling of excitons to coherent optical phonons modulates the intensity and the energy of the excitonic resonance at the phonon frequency on the ps time scale~\cite{Mor2021,sayers2023,Hase2003}. 

While the interaction of excitons with either charge carriers, or other excitons, or phonons have been separately addressed by several spectroscopy works~\cite{Carvalho2015, Ni2017, Raja2018,  Trovatello2020, Chen2020, Li2021, Mor2021,sayers2023,Mor2022,Boschini2024} and theoretical studies~\cite{Giustino2017, Mishra2018, Paleari2019}, the signature of exciton-phonon coupling in combination with photoinduced screening is still unexplored. As both many-body contributions determine the nonequilibrium behavior of photoexcited semiconductors, their interplay should crucially impact on the optoelectronic functionalities of any semiconducting system. Thus, unveiling how the dynamics induced by exciton-phonon coupling manifest in presence or absence of transient changes in the screening of the Coulomb interaction is key both from a fundamental and applications view point.

Here, we address the effect of dynamical screening on the spectroscopic signature of the exciton dynamics coupled to coherent phonons in the layered semiconductor Bismuth tri-iodide (\bii). Transient absorption of \bii\, thin films~\cite{Scholz2018} has reported a composite ultrafast exciton and coherent optical phonon dynamics. Moreover, broadband transient reflectance (TR)  of a \bii\, single crystal~\cite{Mor2021} has proven the coupling of excitons to coherently excited A$_g$ optical phonons at 3.4 THz (113.4~cm$^{-1}$) under photoexcitation to the conduction-band continuum. However, how the dynamical screening by the photoexcited carriers modifies the TR spectrum related to the coupled exciton-phonon dynamics has remained undefined. To this goal, we present a comparative TR spectroscopy study of a \bii\, single crystal (see experimental scheme in~\Fref{Figure 1}(a)) performed under different excitation conditions, i.e. either resonant with the exciton transition or above the optical gap into the conduction-band continuum (termed here as 'off-resonance'), in order to exclude or enable the photoexcitation of quasi-free carriers.

\section{Results and discussion}
The layered crystal structure of \bii\, is reported in~\Fref{Figure 1}~(a). The main A$_g$ Raman mode at 3.4~THz (113.4 cm$^{-1}$) involves mainly out-of-plane stretching of the Bi-I bonds, as represented by the black arrows. The lowest bright excitonic resonance is at 2.00 eV at room temperature~\cite{Gosetti2023} and its wave function localizes within up to three atomic layers, as shown by the calculation (blue shades, modified from Mor \textit{et al.}~\cite{Mor2021}). The steady-state reflectance spectrum of \bii\,is reported as green curve in \Fref{Figure 1}~(b)). It exhibits an isolated peak centered at the excitonic resonance, in agreement with the reflectance of \bii\, at room temperature being mainly determined by the imaginary part of the dielectric function~\cite{Jellison1999}. This observation allows us to discuss the TR of \bii\, in terms of photoinduced changes in the optical absorption. The steady-state absorbance (yellow line in \Fref{Figure 1}(b)) becomes significant at energies above ca. 1.85 eV and exhibits a featureless spectrum. This is due to massive light absorption above the optical gap, which limits the resolution of weaker excitonic signatures in agreement with the literature~\cite{Podraza2013}.

\begin{figure}
\includegraphics[width=0.8\columnwidth]{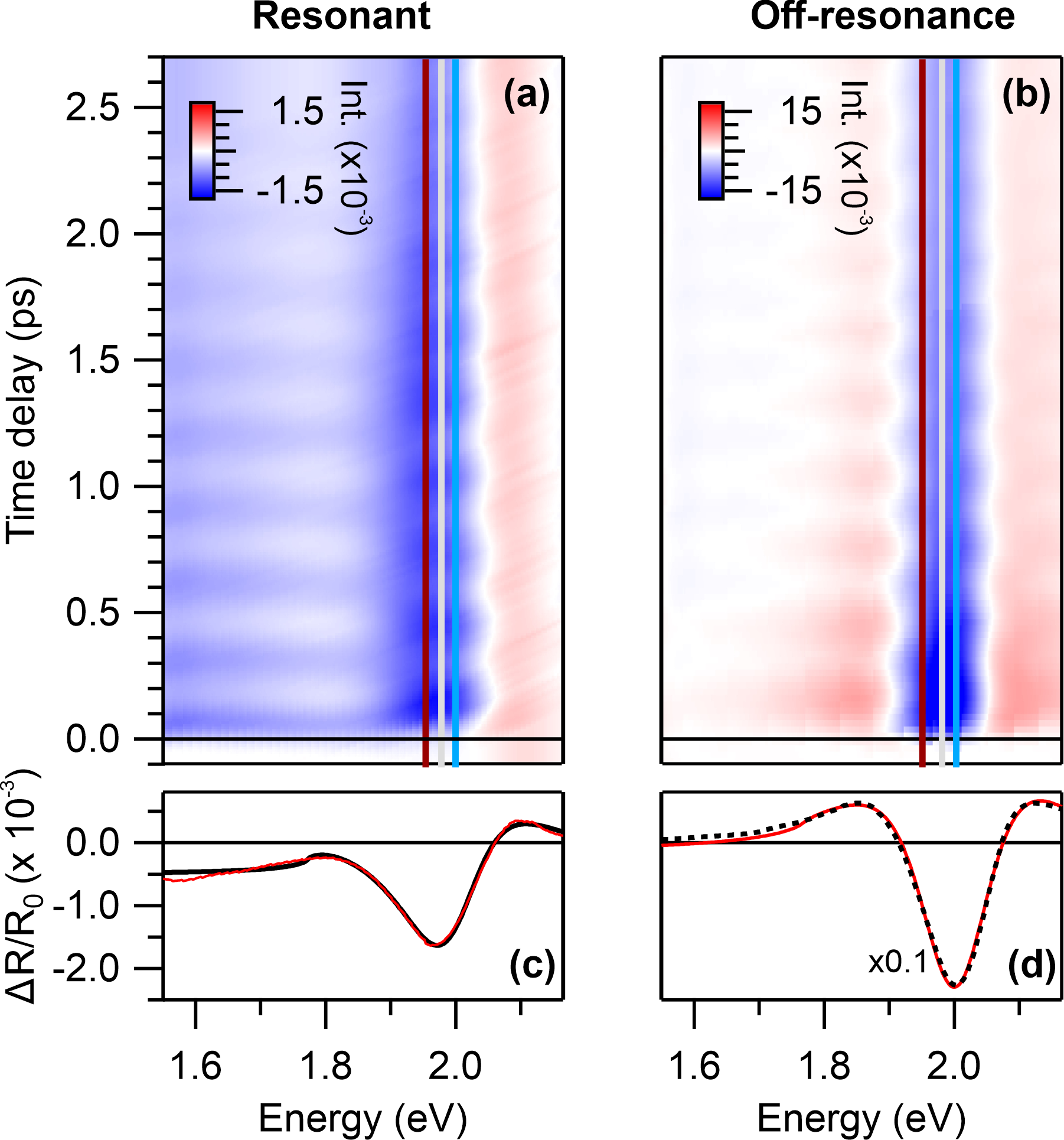}
\caption{$\Delta R /R_0$ intensity maps as a function of pump-probe time delay and probe photon energy for (a) resonant excitation at 1.91~eV and (b) off-resonance excitation at 2.79~eV. The latter data are adapted from Mor \textit{et al.}~\cite{Mor2021} On both maps, the red, gray and blue vertical lines mark the position of the time-resolved line cuts reported in \Fref{Figure 3}. The (c) and (d) panels show the energy-resolved line cuts at $\Delta t$ = 150 fs for the resonant and off-resonance excitation case, respectively.}
\label{Figure 2}
\end{figure}

\Fref{Figure 2} (a) and (b) show the TR intensity $\Delta R/R_0$, i.e. the photoinduced reflectance variation, in false color as a function of pump-probe time delay (left axis) and probe photon energy (bottom axis) under resonant and off-resonance excitation, respectively. Both data are collected at room temperature with absorbed pump fluence of 50~$\mu$J~cm$^{-2}$, which is well within the regime of linear response of the material (see SI for details). For the resonant case, the pump photon energy is tuned well below the conduction band edge in order to avoid the excitation of quasi-free carriers. At the same time, the pump photons excite resonantly to the low-energy side band of the excitonic line (see black triangle in~\Fref{Figure 1}(b)) where the absorbance is still significant. For the off-resonance case, the pump photon energy enables the excitation of electrons to the conduction-band continuum (see white triangle in~\Fref{Figure 1}(b).

$\Delta R/R_0$ under resonant excitation (\Fref{Figure 2}~(a)) shows two spectral regions of negative (blue) and positive (red) intensity  below and above ca. 2.05~eV, respectively. Particularly, an intensity minimum occurs at the exciton energy, and a negative feature extends between ca. 1.55 and 1.80~eV, a region corresponding to the edge of the indirect band gap at 1.67~eV~\cite{Podraza2013}. These features are clearly resolved in the horizontal line cut at 150 fs reported in \Fref{Figure 2}~(c). Moreover, in the TR color map at any probe photon energy, a time-periodic intensity modulation (coherent optical response) superimposes the non-periodic evolution of the TR intensity (incoherent optical response). 
The incoherent optical response reflects the relaxation dynamics of the system towards equilibrium. The coherent optical response is due to the generation of coherent optical phonons by the ultrashort pump pulse~\cite{Pfeifer1992, Dekorsy1999, Hase2004}.  

$\Delta R/R_0$ induced by off-resonance excitation is shown in \Fref{Figure 2}~(b). Again, a coherent optical response adds up to the incoherent one at any energy. The incoherent optical response shows a double sign change with negative intensity centered at the exciton energy, and positive intensity at higher and lower energies. This is also illustrated by the horizontal line cut at 150~fs in \Fref{Figure 2}~(d). All in all, TR reports spectral evidence of the generation of coherent optical phonons as well as excitons under both resonant and off-resonance excitation conditions.

With regard to the incoherent optical response, we assign the negative TR of both data sets to photobleaching of exciton transitions, whereas the positive TR to transient absorption from photoexcited states and photoinduced broadening of the exciton line width. Besides, under off-resonance excitation, free-carrier-induced band gap renormalization is expected to add a positive contribution in the in-gap energy region~\cite{Scholz2018}. Spectral evidence of these band gap dynamics is discussed later as a proof of photoinduced screening. Conversely, under resonant excitation, these transient band gap dynamics should be marginal due to the lack of a population of photoexcited quasi-free carriers. Consistently, the TR is negative at all energies below the exciton peak, suggesting the optical response in that region being rather dominated by photobleaching of phonon-assisted optical transitions across the indirect band gap. 

\begin{figure}
\includegraphics[width=0.9\columnwidth]{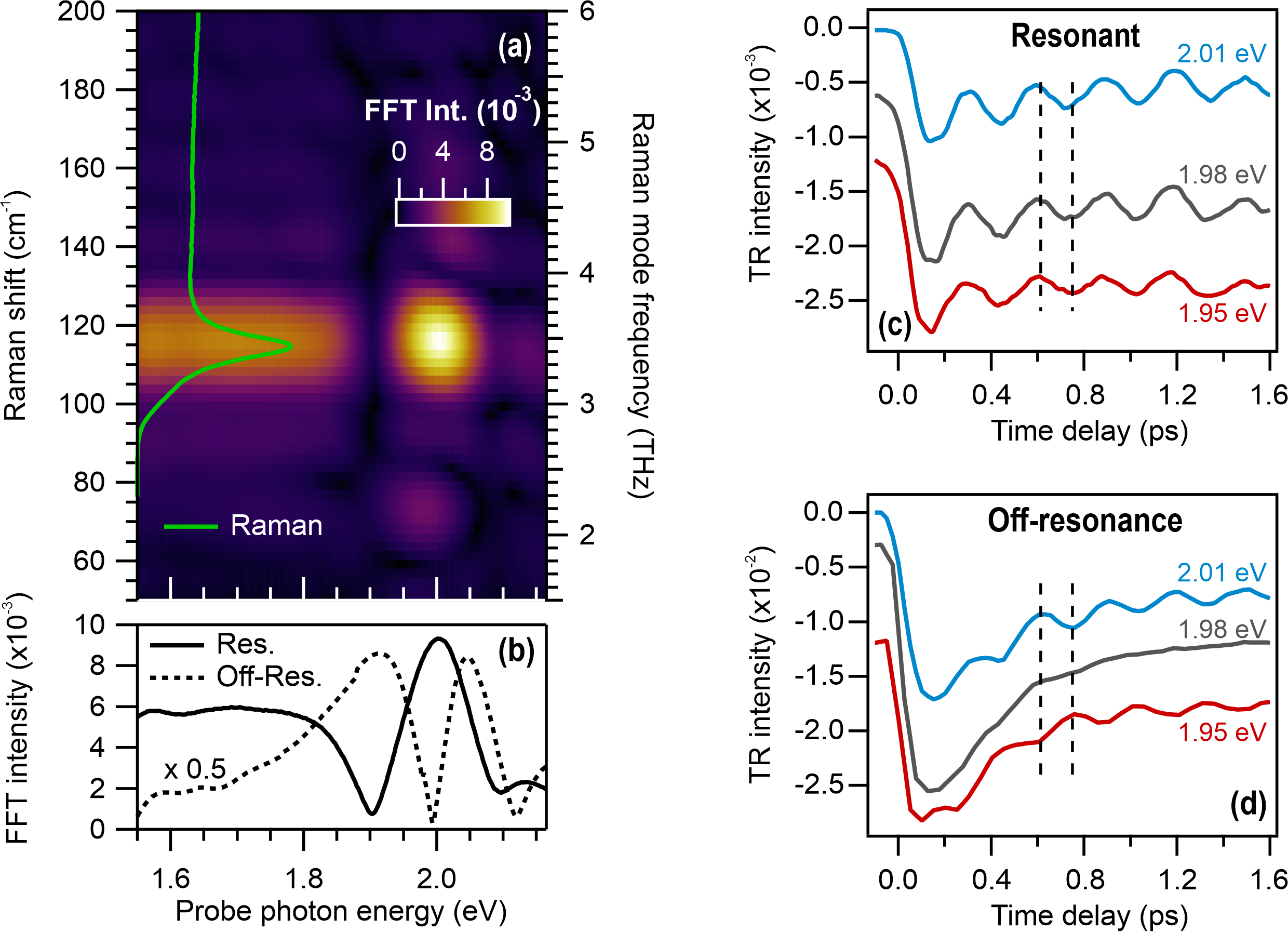}
\caption{(a) Color map of the Fourier transformation of the coherent optical response as a function of Raman shift/mode frequency (left/right axis) for each probe photon energy (bottom axis). The Raman spectrum is shown as green line for reference. (b) Horizontal line cut at the $A_g$-mode frequency for resonant (solid black) and off-resonance (dashed black) excitation. (c) and (d) $\Delta R /R_0$ line cuts at 2.01 eV (blue), 1.98 eV (gray) and 1.95 eV (red) as a function of time delay for resonant and off-resonance photoexcitation, respectively.}
\label{Figure 3}
\end{figure}

We now focus on the coherent optical response. To this goal, we have extracted the bare periodic modulation of the  $\Delta R/R_0$ intensity map upon subtraction of the incoherent TR contribution. The latter is obtained by applying a smoothing filter to the  $\Delta R/R_0$ intensity at each probe photon energy in the window of positive time delays. In \Fref{Figure 3} (a), the Fourier transformation (FT) of the coherent optical response for the resonant case is shown in false color as a function of probe photon energy. We find a single FT component at 3.4~THz (113.4~cm$^{-1}$) which corresponds to the $A_g$ vibrational mode, as confirmed by the Raman spectrum (green curve) overlaid on the FT map. In \Fref{Figure 3}~(b), the FT intensity at 3.4~THz is reported as solid black curve, while the dashed black curve is the corresponding line cut for the off-resonance case (For the relevant FT map, we refer to Mor \textit{et al.}~\cite{Mor2021}). We observe that for resonant photoexcitation, the coherent phonon amplitude at the exciton energy shows a maximum, whereas under off-resonance excitation, a node which was proven to result from a periodic shifting of the excitonic resonance due to coupling to coherent optical phonons~\cite{Mor2021}. Interestingly, such node resembles an antiresonance (dip) obtained from continuous wavelet transformation of the coherent TR of silicon, which was analogously explained by a Fano resonance in the Raman spectrum due to coupling of an LO phonon with the exciton continuum~\cite{Hase2003}. The difference in the photon energy-resolved FT observed in \Fref{Figure 3}~(b) is the first sign that a periodic modulation of the exciton peak energy is hindered under resonant excitation. More generally, it shows that the coupled exciton-phonon dynamics and relevant TR response are affected by the photoexcitation condition.

Further evidence of this finding is found in the time domain. \Fref{Figure 3} (c) shows, for the resonant case, the TR at the exciton energy (gray) and at the low- (blue) and high-energy (red) side of the exciton peak, respectively. The line cuts are extracted from the TR intensity map at the energies marked in \Fref{Figure 2}(a) by vertical lines of the respective colors. All line cuts exhibit an in-phase oscillation persisting for a few ps. Conversely, the intensity oscillation for the off-resonant case shows opposite phase at the low- and high-energy side of the exciton peak, and is completely damped at the exciton peak energy (see  \Fref{Figure 3} (d)). From Mor \textit{et al.}~\cite{Mor2021}, we know that such $\pi$-phase shift of the TR intensity at the exciton energy is connected to the node of the FT intensity, and further identifies the coherent phonon-mediated modulation of the energy of the excitonic resonance. The absence of this other fingerprint corroborates that these dynamics induced by exciton-phonon coupling are suppressed under resonant excitation condition.

\begin{figure}
\includegraphics[width=0.9\columnwidth]
{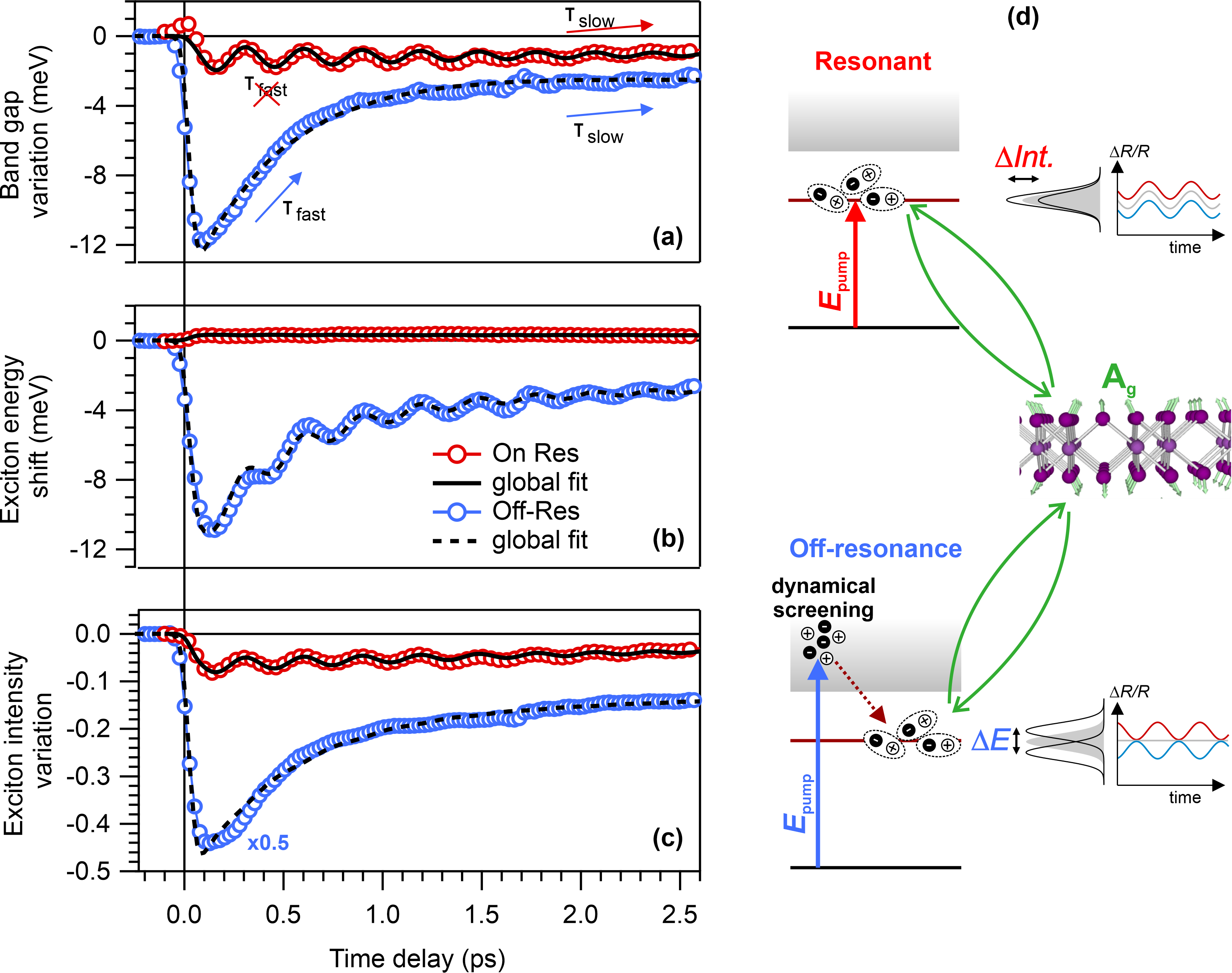}
\caption{Time-resolved variation of (a) the band gap, and of (b) the energy shift and (c) the intensity of the excitonic resonance under resonant (red) and off-resonance (blue) photoexcitation. The black lines are the fits. (d) Schematic of the spectroscopic signature of exciton-phonon coupling in either absence (top) or presence(bottom) of photoinduced screening.
}
\label{Figure 4}
\end{figure}
We now investigate the mechanism underlying the inhibition of a coherent-phonon mediated modulation of the excitonic resonance in the resonant case. We perform differential fitting analysis of the TR spectra at various pump-probe time delays and for both photoexcitation regimes (see red curves in \Fref{Figure 2}(b) and (c)). The analysis consists of two steps. First, we find the best parametrization of the equilibrium reflectance spectrum through Tauc-Lorenz model. Each Tauc-Lorentz oscillator is defined by three parameters: the peak amplitude, the peak energy and the band gap energy. The latter accounts for a non-constant background due to the partially overlapping tail of the upper band gap edge. The result is shown in \Fref{Figure 1}(b) as black curve on top of the equilibrium reflectance data. Then, we fit each TR spectra through variation of the least number of parameters. The fit function is given by $(R_{noneq}-R_{eq})/R_{eq}$, where the best-fit parameters for the equilibrium spectrum, $R_{eq}$, are kept hold in the fitting procedure, while we allow the parameters for the nonequilbrium specturm, $R_{eq}$, to adjust in odrer to fit the TR spectra. As a result, we obtain the variation in the exciton energy and intensity from the change in the Tauc-Lorentz peak energy and amplitude, and in the band gap from the difference with respect to the upper band gap edge at equilibrium. \Fref{Figure 4} shows the differential fit results under both resonant (red) and off-resonance (blue) excitation, i.e. (a) the variation of the band gap, and (b) the energy shift and (c) the intensity variation of the Tauc-Lorenz oscillator accounting for the exciton transition at 2.00~eV. The temporal evolution of those observables is obtained upon \textit{global} fitting with the sum of a double exponential decaying function and an exponentially damped cosine function, convoluted with a Gaussian function that accounts for the pulses' cross correlation. The best fits are shown as black curves on top of the respective data for the resonant (solid) and off-resonance (dashed) case.

Under off-resonance excitation (blue markers), the band gap is significantly reduced by ca. 12~meV at the excitation time, and the recovery dynamics report two timescales, $\tau_{\text{fast}}~=~$473~$\pm$~15~fs and $\tau_{\text{slow}}~>$~2.5~ps. The exciton energy shows an initial shift of ca. 11~meV to \textit{lower} energies, and an intense periodic modulation at the A$_g$ phonon frequency with initial amplitude of 1.2~$\pm$~0.1~meV. Eventually, the exciton intensity is strongly suppressed upon photoexcitation, and recovers on the two time scales $\tau_{\text{fast}}$ and $\tau_{\text{slow}}$ without periodic oscillation at the A$_g$ phonon frequency. Note that the data is scaled by a factor of 0.5 for comparison with the resonant case. The abrupt band gap renormalization attests the activation of dynamical screening due to the excitation of quasi-free carriers to the continuum. The fast relaxation is accordingly assigned to carrier-carrier scattering events. The non-periodic evolution of the exciton (both energy and intensity) mirrors the band gap behavior, indicating that the two relaxation dynamics are connected. Finally, the periodic modulation of the exciton energy signifies once more the coupling to coherent optical phonons in accordance with previous ab-initio calculations~\cite{Mor2021}.

The results under resonant excitation (red markers) are drastically different: (a) the band gap narrows by only 1.5~meV and oscillates at the frequency of the $A_g$ phonon mode, (b) the exciton energy barely reports an almost detection-limited shift by ca. 0.3~meV to \textit{higher} energies at timezero and does not oscillate, and (c) the exciton intensity variation shows an abrupt suppression followed by a recovery on the single-exponential dynamics $\tau_{\text{slow}}$ and a periodic oscillation. Now, the marginal band gap shrinking and the missing fast recovery $\tau_{\text{fast}}$ is consistent with the lack of quasi-free carrier-induced dynamical screening under resonant excitation. Notably, the small blueshift of the excitonic resonance at the excitation time points towards a repulsive interaction among excitons and it is opposite to the shifting direction of the band gap. Eventually, the coupling of excitons to coherent optical phonons manifests \textit{solely} as oscillations of the intensity (which appears also in the TR intensity at any other energy), while it does not trigger any modulation of the exciton energy.

We summarize the effect of exciton-phonon coupling and photoinduced screening on the TR spectrum in \Fref{Figure 4}(d). When excitons and coherent phonons are generated in absence of photoexcited quasi-free carriers (top of \Fref{Figure 4}(d)), the \textit{un}screened exciton-exciton interaction likely govern the dynamics at timezero leading to the initial blueshift of the exciton energy. Consequently, (1) a photo-induced modulation of the excitonic resonance is suppressed and (2) the coupling appears in a modulation of the exciton \textit{intensity}. Conversely, the activation of dynamical screening by photoexcited carriers (bottom of \Fref{Figure 4}(d)) plausibly weakens the repulsion among excitons, while locking the exciton energy modulation to the A$_g$ coherent optical phonons. As a result, the \textit{energy} of the excitonic resonance oscillates in addition to following rigidly the massive renormalization of the band gap. 

\section{Conclusions}
In summary, we succeed to experimentally disentangle the distinct contributions of exciton-exciton repulsion, exciton-carriers interaction and exciton-phonon coupling to the TR spectrum a photoexcited layered semiconductor. We find that the photoinduced screening modifies how exciton-phonon coupling manifest in the TR spectrum. Particularly, it promotes the energy modulation of the excitonic resonance which is phase locked to the atomic vibration. This is opposed to the case in which the \textit{un}screened exciton-exciton repulsion sets the shifting dynamics of the excitonic resonance at the excitation time, and exciton-phonon coupling leads solely to an intensity modulation. To the best of our knowledge, a theory is still missing which can comprehensively capture the interplay of all these many-body interactions on the nonequilbrium optical spectrum. Hopefully, this all-optical study will trigger new theoretical efforts which eventually contribute to the race towards the control of coupled dynamics of condensed matter systems on the ultrafast timescales.

\begin{acknowledgement}
S.M., S.P., L.S., and C.G. acknowledge partial support from D.1 grant of the Universit\'a Cattolica del Sacro Cuore. C.G. acknowledge financial support from MIUR through the PRIN 2015 (Prot. 2015C5SEJJ001), PRIN 2017 (Prot. 20172H2SC4\_005) and PRIN 2022 (Prot. 20228YCYY7) programs.
\end{acknowledgement}

\begin{suppinfo}

The Supporting Information is available free of charge at xxx.

Crystal growth; Optical set-up; Data processing for the correction of the  supercontiuum white-ligh temporal
chirp; \bii\, equilibrium reflectance.

\end{suppinfo}

\bibliography{BiI_bib.bib}

\end{document}